\documentclass[conference]{IEEEtran}
\usepackage{cite}
\usepackage{amsmath,amssymb,amsfonts}
\usepackage{algorithmic}
\usepackage{graphicx}
\usepackage{textcomp}
\usepackage{xcolor}
\def\BibTeX{{\rm B\kern-.05em{\sc i\kern-.025em b}\kern-.08em
    T\kern-.1667em\lower.7ex\hbox{E}\kern-.125emX}}
\usepackage{url}
\begin{document}
\title{Lightweight DNN for Full-Band Speech Denoising on Mobile Devices: Exploiting Long and Short Temporal Patterns}
\author{\IEEEauthorblockN{Konstantinos Drossos}
\IEEEauthorblockA{\textit{Nokia Technologies} \\
Espoo, Finland \\
konstaninos.drosos@nokia.com}
\and
\IEEEauthorblockN{Mikko Heikkinen}
\IEEEauthorblockA{\textit{Nokia Technologies} \\
Tampere, Finland \\
mikko.heikkinen@nokia.com}
\and
\IEEEauthorblockN{Paschalis Tsiaflakis}
\IEEEauthorblockA{\textit{Nokia Bell Labs} \\
Antwerp, Belgium \\
paschalis.tsiaflakis@nokia-bell-labs.com}
}
\maketitle
\begin{abstract}
Speech denoising (SD) is an important task of many, if not all, modern signal processing chains used in devices and for everyday-life applications. While there are many published and powerful deep neural network (DNN)-based methods for SD, few are optimized for resource-constrained platforms such as mobile devices. Additionally, most DNN-based methods for SD are not focusing on full-band (FB) signals, i.e. having 48 kHz sampling rate, and/or low latency cases. In this paper we present a causal, low latency, and lightweight DNN-based method for full-band SD, leveraging both short and long temporal patterns. The method is based on a modified UNet architecture employing look-back frames, temporal spanning of convolutional kernels, and recurrent neural networks for exploiting short and long temporal patterns in the signal and estimated denoising mask. The DNN operates on a causal frame-by-frame basis taking as an input the STFT magnitude, utilizes inverted bottlenecks inspired by MobileNet, employs causal instance normalization for channel-wise normalization, and achieves a real-time factor below 0.02 when deployed on a modern mobile phone. The proposed method is evaluated using established speech denoising metrics and publicly available datasets, demonstrating its effectiveness in achieving an (SI-)SDR value that outperforms existing FB and low latency SD methods.
\end{abstract}

\section{Introduction}
Speech denoising (SD) is crucial task in most applications of communications~\cite{pascual2017segan,9747055}. From large, complex systems to small devices with real-time requirements, there is almost always the need for denoising a speech signal. Especially with the proliferation of technological advances and applications, SD is widely and commonly used as the first step of speech signal enhancement in a processing chain. Recent advances in SD show that methods employing computationally complex deep neural networks (DNNs) outperform legacy/traditional digital signal processing (DSP) based methods. For example, the method in~\cite{luo:2019:taslp} yields results that outperform oracle performance, whereas methods in~\cite{ks-restoration} employ DNNs that can restore a speech signal by alleviating multiple degradations like noise, clicking, packet loss, etc. Although such results are impressive, the DNNs in these methods are too computationally complex to be deployed in small devices for real-time usage. Additionally, the speech signals used in the aforementioned methods typically operate at or below 16 kHz sampling rate.

The above two considerations (i.e. complexity and sampling rate) are important and can prove to be restrictive factors for communication applications that are using high-fidelity and full-band audio (i.e. with 48 kHz sampling rate). For example, modern day handheld or mobile devices that want to use modern voice codecs with full-band audio, like 3GPP Immersive Voice and Audio Services ~\cite{multrus2024immersive}, have to find ways to sidestep the above issues. Some recently published SD methods are focusing on the denoinsing/enhancement of speech signals with real-time (or almost real-time) processing and, in some cases, with using full-band speech signals. In~\cite{tiny-unet,shetu:2024:icassp,rong:2024:icassp} are presented SD methods that exhibit reduced computational complexity, with some of them using the well-known UNet architecture~\cite{ronneberger:2015:miccai}.

While the above methods are focusing on the computational complexity issue, they are using signals that operate on 16 Hz sampling rate. But, there are other methods that seem to tackle the complexity issue while using full-band signals, like~\cite{9747055,Valin2017AHD,Valin2020}. Though, some of these methods seem to have inherent latency, as they are commonly using look-ahead frames (LAFs), for example~\cite{9747055}. In the most common case where an SD method is used in a larger signal processing chain, LAFs can significantly impact the total algorithmic/processing delay of the signal processing chain. Thus, using LAFs may be deemed undesirable for many low latency applications. Additionally, some of the existing low-complexity, full-band methods process both phase and magnitude spectrum, for example~\cite{9747055}. Again, when SD is used in a larger signal processing chain, affecting the phase of the input noisy mixture might be unwanted and have side effects to the rest of the processing chain.

In this paper we focus on communication applications using high-fidelity audio and full-band audio (i.e. 48kHz sampling rate) and we present a UNet-variant SD method that prioritizes the three considerations of the computational complexity, very low latency, and full-band speech signals. Our method is causal, operates on a single frame of short-time Fourier transform (STFT) magnitude, employs learning of short and long temporal patterns in the encoding, decoding, and mask prediction processes, is extremely lightweight (with a real-time factor below 0.02), is focused on full-band signals, and is of low latency (no LAFs) and real-time. We evaluate our method using established speech denoising metrics and publicly available datasets, obtaining state-of-the-art (SI-)SDR on full-band datasets. To the best of our knowledge, our method is outperforming existing methods as a full-band SD method, using no LAFs and having very small real-time factor. The rest of the paper is as follows. Section~\ref{sec:method} holds the presentation of the SD method while Section~\ref{sec:evaluation} presents the followed evaluation process. In Section~\ref{sec:results-and-discussion} are the obtained results with their discussion, and Section~\ref{sec:conclusion} concludes the paper.

%
%
% --------------------------------------------------------
% --------------------------------------------------------
%
%

\section{Proposed method}\label{sec:method}
Our method takes as an input a STFT magnitude vector of a full-band noisy speech signal at time-step (or frame) $t$ and with $F$ frequency bands, $|\tilde{\mathbf{X}}|_{t}\in\mathbb{R}_{\geq0}^{F}$, and outputs its denoised version, $|\hat{\mathbf{X}}|_{t}$. To do so, our method exploits short and long temporal patterns in input signal and learned representations by employing different auxiliary outputs at previous time-steps, two CNN-based modules with kernels that span multiple time-steps, and two RNN-based modules that apply recurrence over the time dimension. Additionally, our method is causal and does not exploit any future time-step information nor require access to look-ahead frames.

\begin{figure}[!t]
    \centering
    \includegraphics[width=0.99\linewidth]{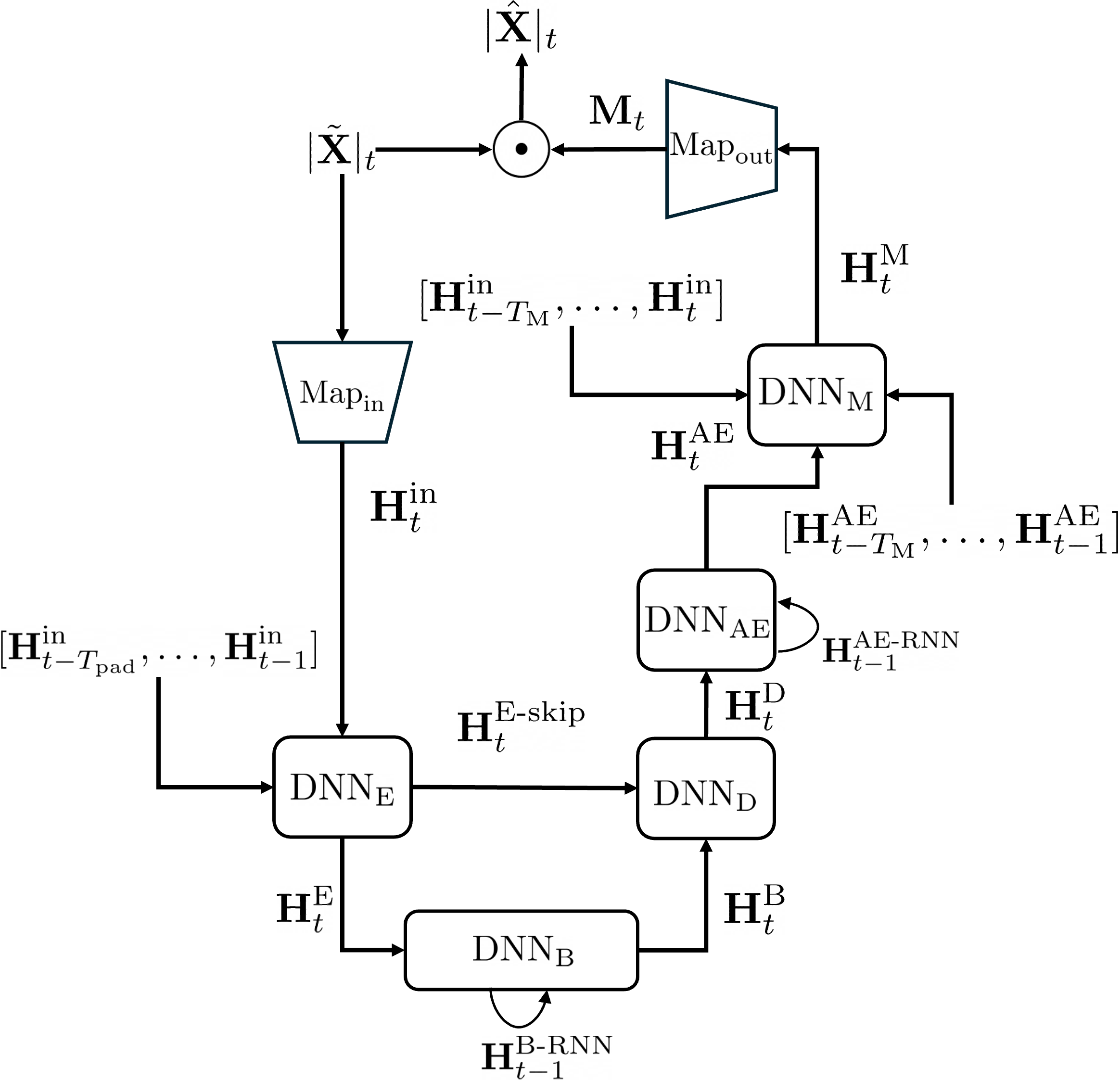}
    \caption{Illustration of the proposed method with the sub-modules. Output matrices' notation is of the form $\mathbf{H}^{L}_{t}$, where $L$ is the corresponding sub-module index/short-hand, seen at the name of the sub-module, $t$ is the time-step index, and $T_{\text{pad}}$ and $T_{\text{M}}$ are the number of previous time-steps used in CNN-based sub-modules. The RNN index signifies specifically the RNN component of sub-modules, and the skip index, skip connections.}
    \label{fig:method}
\end{figure}

There are seven sub-modules in our method, namely: i) a learned mapping/filter-bank $\text{Map}_{\text{in}}$, ii) a CNN-based encoder $\text{DNN}_{\text{E}}$, iii) an RNN-based bottleneck $\text{DNN}_{\text{B}}$, iv) a CNN-based decoder $\text{DNN}_{\text{D}}$, v) an RNN-based autoencoder $\text{DNN}_{\text{AE}}$, vi) a CNN-based mask predictor $\text{DNN}_{\text{M}}$, and vii) a learned output mapping $\text{Map}_{\text{out}}$. The latter outputs a frequency mask at time step $t$, $\mathbf{M}_{t}$. $\mathbf{M}_{t}$ is applied to $|\tilde{\mathbf{X}}|_{t}$, yielding $|\hat{\mathbf{X}}|_{t}$, while the phase of the input noisy signal, $\angle\tilde{\mathbf{X}}_{t}$, is left unaltered and used to synthesize back the time-domain predicted signal from $|\hat{\mathbf{X}}|_{t}$. Figure~\ref{fig:method} illustrates our method and its sub-modules. 

\subsection{Input mapping and encoder}
$\text{Map}_{\text{in}}$ is the first sub-module of our method and is tasked with pre-processing the input vector, normalizing it and reducing its input dimensionality. To do so, $\text{Map}_{\text{in}}$ consists of a learned normalization process and a learnable mapping function, $F \mapsto F'$ with $F > F'$. The mapping function is implemented by a feed-forward neural network (FNN) followed by another normalization process and a nonlinearity. The output of $\text{Map}_{\text{in}}$ is $\mathbf{H}^{\text{in}}_{t}\in\mathbb{R}^{F'}$, obtained as

\begin{equation}
    \mathbf{H}^{\text{map-in}}_{t} = \text{Map}_{\text{in}}(|\tilde{\mathbf{X}}|_{t}).
\end{equation}

\noindent
Our method utilizes the input mapping sub-module as a way to globally reduce the complexity of the method, by enforcing a reduced dimensionality space since the very beginning of the processing. This, effectively, makes the method to operate on its own input dimensionality space, defined by $F'$. 

Next is the CNN-based encoder, $\text{DNN}_{\text{E}}$, designed after the UNet architecture~\cite{ronneberger:2015:miccai}. $\text{DNN}_{\text{E}}$ takes as an input the $\mathbf{H}^{\text{map-in}}_{t}$ and a number of previous outputs from $\text{Map}_{\text{in}}$ equal to $T_{\text{pad}}$, $[\mathbf{H}^{\text{map-in}}_{t-T_{\text{pad}}}, \ldots, \mathbf{H}^{\text{map-in}}_{t-1}]$. The encoder first learns a short-time-pattern informed representation of the current input $\mathbf{H}^{\text{map-in}}_{t}$, and then extracts from the latter an encoded version $\mathbf{H}^{\text{E}}_{t}$, as 

\begin{align}
    \mathbf{H}^{\text{E}}_{t} &= \text{DNN}_{\text{E}}(\mathbf{H}^{\text{in-pad}}),\text{ where}\\
    \mathbf{H}^{\text{in-pad}} &= [(\mathbf{H}^{\text{map-in}}_{t-T_{\text{pad}}})^{\mathrm{T}}; \ldots; (\mathbf{H}^{\text{map-in}}_{t})^{\mathrm{T}}]^{\mathrm{T}}\text{, and}
\end{align}

\noindent
$\mathbf{H}^{\text{E}}_{t}\in\mathbb{R}^{C^{N_{\text{E}}}_{\text{E, out}}\times F^{N_{\text{E}}}}$, 
$\mathbf{H}^{\text{in-pad}}\in\mathbb{R}^{T_{\text{in-pad}}\times F_{\text{in-pad}}}$, and $T_{\text{in-pad}}$ and $F_{\text{in-pad}}$ are the time and frequency related dimensions of $\mathbf{H}^{\text{in-pad}}$. $\text{DNN}_{\text{E}}$ consists of $1+N_{\text{E}}$ concatenated CNN blocks. The first CNN block consists of a 2D CNN, followed by a normalization process and a nonlinearity. The kernel of the 2D CNN at the first CNN block spans $T_{\text{pad}}$ time-steps in the time dimension, outputs $C^{1}_{\text{E, in}} = T_{\text{pad}}$ channels, and the convolution operation has appropriate padding and strides to reduce the time dimension to 1 and not alter the frequency dimension. Essentially, the first CNN block tries to map the short temporal patterns of $\mathbf{H}^{\text{in-pad}}$ in the channel dimension of its output.

The other $N_{\text{E}}$ CNN blocks of the encoder consist of three 1D CNNs, modeled after MobileNet v2~\cite{sandler:2018:cvpr}. The first is a point-wise 1D CNN, with unit kernel, that increases channel dimension of the learned features from $C^{n_{\text{E}}}_{\text{E, in}}$ to $C^{n_{\text{E}}}_{\text{E}}$, the second is a depth-wise 1D CNN, with $K^{n_{\text{E}}}_{\text{E}}$ kernel using $S^{n_{\text{E}}}_{\text{E}}$ stride, that operates on the increased channel/dimensionality space. The third 1D CNN is also point-wise, with unit kernel, that brings down the dimensionality of the features from $C^{n_{\text{E}}}_{\text{E}}$ to $C^{n_{\text{E}}}_{\text{E, out}}$. All three 1D CNNs are followed by a normalization process and only the first two have a nonlinearity following the normalization. 

Our method uses the UNet architecture as the basis for the encoder in order to, on the one-hand, utilize a widely adopted architecture with proven well performance in the speech denoising task and, on the other hand, enable the adoption of complexity reduction techniques that have been developed and applied for CNN-based modules. Regarding the complexity aspect, our method employs the MobileNet v2 structure and the inverted bottlenecks, allowing both effective processing of the learned features and reduced complexity and number of learnable parameters. Additionally, to take advantage of short-time temporal patters, the encoder of our method takes as an input previous outputs of the mapping process, further showing the impact of the mapping process and enhancing the temporal perception of the encoder. 

\subsection{Bottleneck}
$\mathbf{H}^{\text{E}}_{t}$ is given as an input to the bottleneck $\text{DNN}_{\text{B}}$ of our method. $\text{DNN}_{\text{B}}$ process its input, exploiting long temporal patterns between the current and previous inputs, and then outputs $\mathbf{H}^{\text{B}}_{t}$ as

\begin{equation}
    \mathbf{H}^{\text{B}}_{t} = \text{DNN}_{\text{B}}(\mathbf{H}^{\text{E}}_{t})\text{,}
\end{equation}

\noindent
where $\mathbf{H}^{\text{B}}_{t}$ and $\mathbf{H}^{\text{E}}_{t}$ have same dimensionality. $\text{DNN}_{\text{B}}$ consists of a 1D CNN that maps all features to the channel dimension, effectively creating a 1D vector from the 2D (channel and feature dimensions) of its input, an RNN that process the output of the 1D CNN, and then another 1D CNN that maps back the output of the RNN to the same dimensionality of $\mathbf{H}^{\text{E}}_{t}$. Effectively, the bottleneck utilizes the 1D CNNs to map the input of the bottleneck back and forth to a vector, so the RNN can use that vector. The RNN is employed in order to discover and exploit long temporal patterns at the output of the encoder, enhancing both the temporal perception of the whole method and the decoding process. 

Given the above, the input to the first 1D CNN of the bottleneck is shuffled at its dimensions, where the features and channels dimensions are swapped. The 1D CNN applies a unit kernel reducing the number of channels to 1, effectively creating a vector. This vector is given as an input to the RNN of the bottleneck, along with the previous output of the RNN, which is indicated as $\mathbf{H}^{\text{B-RNN}}_{t-1}$ in Figure~\ref{fig:method}. Finally, the last 1D CNN of the bottleneck takes as an input the output of the RNN, restores the channels back to the number of channels before the first 1D and the dimensions are again swapped to match the dimensionality of $\mathbf{H}^{\text{E}}_{t}$. 

\subsection{Decoder}
The decoder of our method, $\text{DNN}_{\text{D}}$, takes as an input the output of the bottleneck and the skip connections from the encoder, $\text{DNN}_{\text{D}}$, and gradually expands the output of bottleneck to match the dimensionality defined by the input mapping process, $\text{Map}_{\text{in}}$, i.e., restore the feature dimension to $F'$, as

\begin{equation}
    \mathbf{H}^{\text{D}}_{t} = \text{DNN}_{\text{D}}(\mathbf{H}^{\text{B}}_{t}, \mathbf{H}^{\text{E-skip}}_{t})\text{,}
\end{equation}

\noindent
where $\mathbf{H}^{\text{D}}_{t}\in\mathbb{R}^{F'}$, $\mathbf{H}^{\text{E-skip}}_{t}$ are the skip connections from the encoder, as also shown in Figure~\ref{fig:method}, and the concatenation of the skip connection with the input to each CNN block is performed at the channel dimension. $\text{DNN}_{\text{D}}$ is also based on UNet architecture, employing a series of $N_{\text{D}}=N_{\text{E}}$ CNN blocks based on transposed convolutions and skip connections with the corresponding blocks of the encoder, $\text{DNN}_{\text{E}}$. Obtaining and handling the skip connections is according to the UNet model~\cite{ronneberger:2015:miccai}. 

Each of the CNN blocks of the decoder consists of one 1D CNN and one 1D transposed CNN, each of those followed by a normalization process and a nonlinearity. The first CNN maps extract from its inputs (the previous layer and the skip connection) learned features and then the second CNN expands the frequency dimension to match the input of the corresponding encoder CNN block, like in the typical UNet architecture/model. 

\subsection{Output processing}
The output of the decoder, $\mathbf{H}^{\text{D}}_{t}$ is further processed by the RNN-based autoencoder, $\text{DNN}_{\text{AE}}$, in order to further utilize long temporal patterns. In the bottleneck, the RNN that is employed exploits long temporal patterns at the output of the encoder and thus focusing on enhancing the learning of and mapping to the clean-speech manifold, by the encoder and bottleneck. Though, at the $\text{DNN}_{\text{AE}}$, the RNN is employed to exploit long temporal patterns closer to the mask prediction, enhancing the temporal structure of the features that drive the prediction of the denoising mask.

Given that $\mathbf{H}^{\text{D}}_{t}$ is a vector, $\text{DNN}_{\text{AE}}$ consists of a FNN that takes as an input $\mathbf{H}^{\text{D}}_{t}$ and reduces its dimensionality, an RNN that takes as an input the output of the first FNN, and another FNN that increases the dimensionality of the output vector from the RNN, back to $F'$. Both FNNs are followed by a normalization process and the first FNN also includes a nonlinearity after the normalization. The output of the $\text{DNN}_{\text{AE}}$, is

\begin{equation}
    \mathbf{H}^{\text{AE}}_{t} = \text{DNN}_{\text{AE}}(\mathbf{H}^{\text{D}}_{t})\text{, }
\end{equation}

\noindent
where $\mathbf{H}^{\text{AE}}_{t}\in\mathbb{R}^{F'}$. 

$\mathbf{H}^{\text{AE}}_{t}$ is then given as an input to a further 2D CNN, the $\text{DNN}_{\text{M}}$, along with $T_{\text{M}}$ vectors of the mapped noisy magnitude spectrogram, $[\mathbf{H}^{\text{M}}_{t-T_{\text{M}}},\ldots, \mathbf{H}^{\text{M}}_{t}]$, and previous outputs of the $\text{DNN}_{\text{AE}}$, $[\mathbf{H}^{\text{AE}}_{t-T_{\text{M}}},\ldots, \mathbf{H}^{\text{AE}}_{t-1}]$, as

\begin{align}
    \mathbf{H}^{\text{M}}_{t} &= \text{DNN}_{\text{M}}(\mathbf{H}^{\text{AE}}, \mathbf{H}^{\text{in}})\text{, where}\\
    \mathbf{H}^{\text{in}} &= [\mathbf{H}^{\text{in}}_{t-T_{\text{M}}},\ldots,\mathbf{H}^{\text{in}}_{t}]\text{,}\\
    \mathbf{H}^{\text{AE}} &= [\mathbf{H}^{\text{AE}}_{t-T_{\text{M}}},\ldots,\mathbf{H}^{\text{AE}}_{t}]\text{, and}
\end{align}

\noindent
$\mathbf{H}^{\text{AE}}$ and $\mathbf{H}^{\text{in}}$ are concatenated at the channel dimension. Effectively, $\text{DNN}_{\text{M}}$ is tasked with predicting the denoising mask in the reduced dimensionality of $F'$ features, using current and $T_{\text{M}}$ previous vectors of the mapped noisy input spectrogram, $\mathbf{H}^{\text{in}}$, and the output of the $\text{DNN}_{\text{AE}}$. Thus, exploiting short temporal patterns in the correspondence of the noisy input and the RNN-based autoencoder output. Must be noted that there is no normalization or nonlinearity following the $\text{DNN}_{\text{M}}$. 

\subsection{Denoising mask prediction and application}
The denoising mask, $\mathbf{M}_{t}$, is predicted by $\text{Map}_{\text{out}}$ as

\begin{equation}
    \mathbf{M}_{t} = \text{Map}_{\text{out}}(\mathbf{H}^{\text{M}}_{t})\text{,}
\end{equation}

\noindent
where $\text{Map}_{\text{out}}$ is a learnable mapping function, $F' \mapsto F$, mapping the output of $\text{DNN}_{\text{M}}$ back to the dimensionality of $|\tilde{\mathbf{X}}|_{t}$. $\text{Map}_{\text{out}}$ is implemented by an FNN, followed by a sigmoid function that constricts the output of the FNN at the $[0, 1]$ range and, thus, ensuring the value range of $\mathbf{M}_{t}$.

Finally, the predicted magnitude spectrogram of the denoised speech signal is obtained by

\begin{equation}
    |\hat{\mathbf{X}}|_{t} = |\tilde{\mathbf{X}}|_{t} \odot \mathbf{M}_{t}\text{,}
\end{equation}

\noindent
where $\odot$ is the Hadamard (element-wise) product. 

%
%
% --------------------------------------------------------
% --------------------------------------------------------
%
%

\section{Evaluation}\label{sec:evaluation}
For optimizing and evaluating our method we use the publicly available datasets from the DNS-Challenge~\cite{dubey:2023:icassp} and VCTK test set~\cite{vctk,valentini:2016:vctk}, respectively. We assess the performance using typical speech-denoising metrics aligned with existing published methods, and we do model selection and parameter optimization using a validation split of the training data along with the early stopping policy. 

\begin{table*}[!ht]
\centering
\caption{Results of our method (Ours), PercepNet (PNet), DeepFilter v1 (DFv1), and v3 (DFv3). Value of ``-'' means that the corresponding info is not available.}
\begin{tabular}{l|ccccccc}
         & \textbf{SD-SDR} & \textbf{SI-SDR}& \textbf{STOI} & \textbf{PESQ-WB}& \textbf{MSIG} & \textbf{MOVRL}  \\
      \hline\\
PercepNet&   -   &         -      &      -        &  2.54         &        -      &   -           \\
DFv1     &   -   &        16.63   & \textbf{0.94} &  2.81         &      4.14     & 3.46          \\
DFv3     &   -   &         -      & \textbf{0.94} & \textbf{3.17} & \textbf{4.34} & \textbf{3.77} \\      
Ours     & 24.64 & \textbf{22.34} & \textbf{0.94} &  2.82         &      3.58     & 3.29          \\
\end{tabular}
\label{tab:results}
\end{table*}

\subsection{Data and data pre-processing}
We use the DNS-challenge dataset~\cite{dubey:2023:icassp} (referred to as dev-set from now on) as our training and validation data, and the VCTK test set~\cite{valentini:2016:vctk}, which is a subset of the complete VCTK dataset~\cite{vctk} and referred to as test-set from now on, as our testing data since it is the only clean-speech and freely-available testing dataset with 48kHz sampling rate. Our dev-set (i.e., the DNS-challenge dataset) consists of several clean-speech datasets, including the complete VCTK dataset, and the different noise datasets. The test-set is a subset of the VCTK dataset that is used for full-band systems testing in many previous work, e.g.~\cite{9747055,Valin2020}.

For having a fair testing process, we excluded any overlapping clean speech files of VCTK dataset between the test-set and the dev-set. Furthermore, to ensure an appropriate level of quality for the dev-set, we employed the DNSMOS~\cite{reddy2021dnsmos} and filtered the dev-set by removing all clean speech files that had $\text{MOS-SIG} < 4.0$ and $\text{MOS-OVRL} < 3.9$. Then, we randomly split the remaining data in the dev-set to training and validation splits, using a ratio of 80-20\%, respectively, for both clean-speech and noise. Finally, we resampled as necessary the files in the dev-set to 48 kHz sampling rate and split all clean speech and noise files to non-overlapping segments of four seconds duration, with padding at the beginning if needed. 

For the input to our method, we sample each clean speech signal and mixed it maximum two noise files, with an integer signal-to-noise ratio value uniformly sampled from the range of $-10$ to $25$, obtaining $\tilde{\mathbf{X}}$. Then, we scale the absolute peak amplitude of $\tilde{\mathbf{X}}$ to a randomly sampled value between $0.001$ to $0.999$, and we apply STFT with 2048 points ($\approx40$ms) and 50\% overlap, obtaining the magnitude spectrogram and phase, $|\tilde{\mathbf{X}}|$ and $\angle\tilde{\mathbf{X}}$, respectively, both sequences of $T$ vectors with $F$ elements each.

\begin{table}[!ht]
\centering
\caption{$N_{\text{params}}$,  MACS, and RTF of $\text{Ours}_{\text{w/}}$, $\text{Ours}_{\text{w/o}}$, PNet, and DFv1 (no available values for DFv3). Must be noted that the RTFs are obtained on different devices and the presented numbers might be pessimistic for our method, given the hardware used for calculating the corresponding RTFs.}
\begin{tabular}{l|ccc}
                          &$\mathbf{N_{\text{params}}}$ (M) & \textbf{MACS} (G) & \textbf{RTF}\\
  \hline\\
$\text{Ours}_{\text{w/}}$ &       0.451                   & 0.0064              & 0.014\\
$\text{Ours}_{\text{w/o}}$&       \textbf{0.253}          & \textbf{0.0061}     & \textbf{0.012}\\
PNet                      &       8.000                   & 0.8000              &   -  \\
DFv1                      &       1.780 (or 2.310)        & 0.3500 (or 0.3600)  & 0.040
\end{tabular}
\label{tab:results-rtf}
\end{table}

\subsection{Method hyper-parameters and training}
Our method is implemented having hard-swish~\cite{mobilenetv3} as all nonlinearities and causal instance norm as the choice for all normalization processes, apart $\text{Map}_{\text{out}}$ where a sigmoid is used. The reason for the former is to reduce the computational complexity and for the latter the observed performance improvement on the validation split. We use $F'=96$ elements as size of the mapping, $T_{\text{pad}} = 32$ and $T_{\text{M}} = 3$ elements as the look-back for learning short-term structures in the input pre-processing and output processing of the mask, and $N_{\text{E}} = N_{\text{D}} = 6$ encoder and decoder blocks, respectively. It has to be noted that our method uses no look-ahead frames, significantly differentiating our method from others that are using look-ahead frames and thus inducing extra latency equal to the number of look-ahead frames. 

We used $C^{n_{\text{E}}}_{\text{E}} = 256$ as the output channels of the first CNN and $C^{n_{\text{E}}}_{\text{E, out}} = 16$ as the output channels of the third CNN in each encoder block, except the $C^{N_{\text{E}}}_{\text{E, out}} = 64$. 
Kernels, $K^{n_{\text{E}}}_{\text{E}}$, were 5, 3, 5, 3, 5, and 3, for $n_{\text{E}} = 1,\ldots,6$, with corresponding strides $S^{n_{\text{E}}}_{\text{E}}$ 2, 1, 2, 1, 2, and 2, and proper padding for halving the features with stride equals 2. Input and output channels for the second CNN of the decoder are 64 and the kernels and the strides of the second CNN in the decoder are symmetric to the encoder.  All RNNs of our method are gated recurrent units (GRUs), having only one layer.\cite{gru}

We do training for our method using the training split of our dev-set and randomly sampling all $|\tilde{\mathbf{X}}|$, creating batches of 32 sequences. We utilize $\angle\tilde{\mathbf{X}}$ along with the $|\hat{\mathbf{X}}|$ in order to synthesize $\hat{\mathbf{X}}$ and use SI-SDR between $\hat{\mathbf{X}}$ and $\mathbf{X}$ as a loss function. We employ the early stopping policy using the improvement of the validation loss, with a patience of 200 epochs. We use Adam optimizer~\cite{adam} with an initial learning rate of $1e^{-4}$ and default values for the other hyper-parameters, reducing further the learning rate when the validation loss was not improving for 10 consecutive epochs. Additionally, observing the norm of the gradients and the loss curves during training, we clipped the gradients of the parameters to have a $2-$norm of maximum 0.5. 

\subsection{Metrics and baselines}
We assess the performance of our method by the signal-to-distortion (SDR) ratio, both the scale-dependent (SD-SDR) and scale-invariant (SI-SDR)~\cite{sisdr} versions, the short-time-objective-intelligibility (STOI)~\cite{stoi}, the wide-band perceptual evaluation of speech quality (PESQ-WB)~\cite{pesq}, and the DNSMOS outputs of signal (MOS-SIG) and overall (MOS-OVRL)~\cite{reddy2021dnsmos}. Also, to assess the computational complexity of our method, we calculate the average inference time over 100 forward passes. To do so, we exported our model to TFLite format and used the benchmark tool provided by TFLite, on a Pixel 7 phone, using its CPU and no accelerators. From the inference time we calculate the real-time factor for our method. Since other methods use hard-coded features (like ERB) while our method uses a learned one, and for a fair comparison, we provide the real-time factor of our method with and without the input and output mapping processes. We compare our obtained results against the previous, full-band, speech-denoising methods PercepNet~\cite{Valin2020}, DeepFilterNet (DF) v1~\cite{9747055}, and DF v3~\cite{df3}. 

%
%
% --------------------------------------------------------
% --------------------------------------------------------
%
%

\section{Results and discussion}\label{sec:results-and-discussion}
In Table~\ref{tab:results} are the obtained metrics for our method and other existing methods on full-band speech denoising, employing SD- and SI-SDR, STOI, PESQ-WB, DNSMOS-SIG (MSIG), and DNSMOS-OVRL (MOVRL). Also, in Table~\ref{tab:results-rtf} are numbers of parameters ($N_{\text{params}}$), multiply-accumulate operations per second (MACS), and the obtained real-time factors (RTFs) of our method with and without the input/output mapping, Ours\textsubscript{w/} and Ours\textsubscript{w/o}, respectively, and the others methods included in Table~\ref{tab:results}. RTFs of the others methods in Table~\ref{tab:results-rtf} are after~\cite{df2} and are calculated on a notebook while ours on a mobile phone.

As can be seen from Table~\ref{tab:results}, our method provides the highest SI-SDR with a difference of over 6 dB, although it has the lowest complexity and latency of all, meaning that our method does not employ any look-ahead frames, and is the one with the smallest MACs and RTF (as seen in Table~\ref{tab:results-rtf}). At the same time, our method is on par with the other methods regarding STOI and also surpasses all but one other methods on PESQ-WB. The only other method that surpasses our own at the PESQ-WB is the third version of DF. The lower PESQ-WB score of our method, despite better SI-SDR and equivalent STOI, is likely because PESQ-WB only evaluates up to 8 kHz, potentially missing enhancements our model makes in the 8 to 24 kHz region.

In contrast, DFv.3 has utilized considerable focus on processing the frequency regions of the input signal that are valid for a sampling rate of 16 kHz, aligning better with the PESQ evaluation range. Additionally, PESQ-WB penalizes spectral and temporal artifacts that may arise from our fullband processing, even if they are perceptually negligible. These findings, and given the high STOI values, suggest that our method achieves a better overall signal reconstruction (as shown by higher SI-SDR), without compromising intelligibility, even if PESQ-WB is conservative in scoring fullband methods. The above could also explain the discrepancies between the DNSMOS values, between our method and the compared ones. Though, this is just a hypothesis and further exploration is needed. 

Regarding complexity, and as can be seen from Table~\ref{tab:results-rtf}, our method is significantly less complex and faster. Specifically, our method exhibits less MACS by a significant factor of $10^{2}$ while it is almost as four times faster, even with the learned input/output mapping and with the RTF measured on a mobile phone. Even though the utilized devices for measuring the RTFs are not the same, both our and the device used for the other methods in Table~\ref{tab:results-rtf}, use single threaded execution and no accelerators. Additionally, our method is evaluated on a mobile phone, while the others methods in Table~\ref{tab:results-rtf} on a laptop. Finally, the DNN of our method without the input and output mapping (Ours\textsubscript{w/o}) yields $\approx14\%$ smaller RTF compared with Ours\textsubscript{w/}, highlighting the impact of the mapping, either learned or hard-coded (like ERB), to the RTF of the SD, low-latency methods. 

%
%
% --------------------------------------------------------
% --------------------------------------------------------
%
%

\section{Conclusion}\label{sec:conclusion}
In this paper is presented a real-time (with extreme small real-time factor), low latency (without look-ahead frames) DNN-based method for full-band speech denoising. The method is according to the UNet architecture, employing look-back frames and RNNs to model short and long-term temporal patterns, in both the latent space and the output mask. The method outperforms existing methods in SDR related metrics. It performs relatively less good on metrics like PESQ-WB and DNSMOS, but we conjecture that this is because these latter measures are tailored mainly to the 16 kHz range, whereas our proposed method is developed for a 48 kHz range. Finally, the presented method appears to be significantly faster and less complex, having almost four times smaller real-time factor and less MACS by a factor of around $10^{2}$, and thus making the method particularly well-suited for deployment on mobile and resource-constrained devices.

%
%
% --------------------------------------------------------
% --------------------------------------------------------
%
%

\bibliographystyle{IEEEtran}
\bibliography{IEEEabrv,refs}
\end{document}